\documentclass[pdflatex,sn-basic]{sn-jnl}

\usepackage{amsmath}
\usepackage{amssymb}
\usepackage{bm}
\usepackage{booktabs}
\usepackage{multirow}
\usepackage{graphicx}
\usepackage{longtable}
\usepackage{array}
\usepackage{url}
\usepackage{natbib}

\usepackage{tikz}
\usepackage{pgfplots}
\usepgfplotslibrary{statistics}

\usepackage{pgfplotstable}
\usetikzlibrary{arrows.meta,positioning,shapes.geometric}
\pgfplotsset{compat=1.18}

\usepackage{tikz}
\usetikzlibrary{fit,positioning}

\begin{document}

\title[Privacy-Aware Synthetic Data Generation]{
An Energy-Driven Framework For
Privacy-Aware Synthetic Data Generation
}

\author*[1]{\fnm{P.} \sur{Massoli}}
\email{pimassol@istat.it}

\author[2]{\fnm{F.} \sur{Spagnuolo}}
\email{spagnuol@istat.it}

\affil[1]{
\orgdiv{Directorate for Methodology and Statistical Process Design},
\orgname{Italian National Institute of Statistics (ISTAT)},
\orgaddress{
\street{Via Cesare Balbo 16},
\city{Rome},
\postcode{00184},
\country{Italy}
}
}

\affil[2]{
\orgdiv{Division for Development and Enhancement of the Integrated
Registers System by Theme},
\orgname{Italian National Institute of Statistics (ISTAT)},
\orgaddress{
\street{Via Cesare Balbo 16},
\city{Rome},
\postcode{00184},
\country{Italy}
}
}


\abstract{
Access to microdata in official statistics requires balancing
disclosure risk and statistical utility.
This paper proposes an interpretable, open-box energy-driven
framework for privacy-aware synthesis of mixed-type tabular
data, combining discriminative modelling, Bayesian-network
proposals, Metropolis--Hastings sampling and post-generation
optimization.
Explicit energy components govern record-level plausibility,
proximity to training records and diversity, while target
selection balances fidelity and classification utility,
making these trade-offs directly inspectable.
Evaluation uses an artificially constructed individual-level
dataset with demographic, behavioural and health-related
variables.
Validation combines marginal, bivariate and conditional
fidelity diagnostics, diversity measures, nearest-neighbour
analyses and membership inference attacks.
An arrangement of classifiers trained on reference and synthetic data
is evaluated on a common reference holdout.
Repeated target selection assesses stability within
a fixed candidate pool.
Results document the fidelity--utility trade-off obtained under the
evaluated generation and selection settings.
Membership attacks perform close to chance across repeated
splits, indicating limited training-membership discrimination
under the evaluated protocol.
The findings support the feasibility of balancing fidelity,
utility and privacy-aware objectives through explicit
generation and selection criteria, without implying formal
privacy guarantees.
}
	
\keywords{Synthetic Microdata, Privacy-Aware Data Synthesis,
Interpretable Generative Models, Energy-Driven Sampling,
Fidelity--Utility Testing, Official Statistics}

\maketitle


\section{Introduction}
\label{sec:introduction}

The problem of generating synthetic records from observed real
microdata has become increasingly relevant in contemporary
data analysis, where data accessibility is restricted by
confidentiality requirements.
In official statistics, synthetic datasets support analytical
workflows while reducing the disclosure risk associated with
disseminating individual-level observations.
An early formulation conceptualized fully synthetic datasets
as a disclosure limitation mechanism rooted in multiple
imputation \citep{rubin1993}.
Subsequent developments extended this framework to partially
synthetic data, wherein selected sensitive variables are
replaced while the remaining information is retained
\citep{reiter2005}.
More recently, deep generative and autoregressive models
have provided flexible approaches to the representation
of complex, high-dimensional distributions
\citep{goodfellow2014,kingma2014,rezende2014,germain2015,oord2016}.
A central requirement of synthetic data generation is
statistical fidelity: synthetic datasets should reproduce
relevant marginal and conditional distributions and
multivariate associations to support downstream analysis
\citep{elemam2020,snoke2018}.
This requirement is particularly challenging for mixed-type
tabular data characterized by nonlinear dependencies,
rare configurations and high-cardinality attributes.
Distributional fidelity and predictive utility are related
yet distinct. Agreement in selected distributions does not
necessarily preserve the performance of a downstream model,
while comparable predictive accuracy does not establish
broader distributional agreement.
These objectives must also be considered alongside disclosure
risk. Predictive and generative models can expose information
about their training observations through membership
inference attacks \citep{shokri2017,salem2019}.
Rare configurations may contribute simultaneously to
predictive relationships and individual identification,
making statistical usefulness and disclosure risk difficult
to separate. Synthetic records therefore do not provide
protection merely by being artificially generated
\citep{stadler2022}.
Differential privacy provides a rigorous mathematical
framework for controlling the influence of individual
records, often through calibrated noise \citep{dwork2014}.
Its effect on utility depends on the privacy budget,
the mechanism and the characteristics of the data.
Alternative privacy-aware approaches can discourage
proximity to real observations through explicit generation
criteria, but their disclosure properties require empirical
assessment and should not be equated with formal privacy
guarantees.
Existing generative approaches offer different representations
of the underlying dependency structure.
Generative Adversarial Networks and Variational Autoencoders
provide flexible models of complex multivariate distributions
\citep{goodfellow2014,kingma2014,rezende2014}, whereas
copula-based methods represent dependencies through
tractable probabilistic constructions
\citep{sklar1959,bedford2002}.
Their application to heterogeneous tabular data involves
choices concerning model flexibility, interpretability
and the representation of sparse configurations.
Diffusion-based approaches, including STaSy
\citep{kim2023stasy}, TabDDPM
\citep{kotelnikov2023tabddpm}, CoDi
\citep{lee2023codi}, TabSyn
\citep{zhang2024tabsyn} and TabDiff
\citep{shi2025tabdiff}, further extend the capacity
to approximate mixed-type distributions.
However, distribution learning does not necessarily expose
privacy, diversity and task-specific utility as separate,
directly controllable objectives.
Their empirical assessment remains necessary even when
the fitted generator reproduces broad distributional features.
Here, an energy is a scalar penalty that assigns lower values to
configurations better satisfying specified criteria. In a probabilistic
energy-based formulation, lower energy corresponds to greater probability
through an exponential transformation \citep{lecun2006}. This provides
a direct way to express competing requirements and inspect their effects.
The proposed framework uses energy at two levels: penalties guide the
generation of individual records, and a separate objective guides the
selection of a complete synthetic dataset. The fitted statistical models
supply inputs to these criteria; they are not jointly trained as a single
energy-based neural model.

This study proposes an interpretable, open-box formulation
that distinguishes record-level generation criteria from
dataset-level selection objectives.
A pool-construction phase constructs a candidate pool through
Bayesian-network proposals and Metropolis--Hastings sampling,
guided by explicit plausibility, privacy and diversity
penalties.
A target-selection phase selects a fixed-size target through
simulated annealing, balancing distributional fidelity
and predictive utility.
Discriminative learning supplies the plausibility criterion
through gradient boosting \citep{chen2016xgboost}.
The contribution lies in the integration of these established
methods into a generation and selection architecture whose
competing objectives are explicitly specified and separately
assessable.
Its open-box character concerns the sampling and optimization
criteria, rather than an assumption that every fitted
learning model is intrinsically transparent.
The empirical study evaluates mixed-type individual-level
data through distributional comparisons, diversity and
proximity diagnostics, membership inference attacks and
external classification experiments.
Repeated target generation assesses stability conditional
on a fixed pool, while sensitivity analysis examines the
utility weight and optimization budget.
This evaluation distinguishes the criteria used to select
synthetic records from the performance of models subsequently
trained on the released data.
The framework is particularly relevant to official statistics,
where reproducibility and explicit assessment of analytical
usefulness and disclosure risk are essential.
Section~\ref{sec:theoretical_background} reviews the
theoretical foundations, and
Section~\ref{sec:proposed_approach} presents the methodology.
Section~\ref{sec:case_study} describes the case study,
while Section~\ref{sec:results} reports the empirical
evaluation.
Section~\ref{sec:sensitivity_analysis} examines sensitivity
to the utility weight and optimization budget.
Sections~\ref{sec:discussion} and~\ref{sec:conclusion}
discuss the findings and provide concluding remarks.


\section{Theoretical Background}
\label{sec:theoretical_background}
This section introduces the concepts underlying the proposed
sampling and optimization framework, together with the
criteria used to assess synthetic data quality.
The sequence follows the workflow: discriminative learning scores candidate
plausibility, a Bayesian network generates candidate records, and
Metropolis--Hastings uses an energy to accept or reject them. Simulated
annealing subsequently selects a fixed-size dataset from the collected
pool. Distributional distances and membership attacks assess different
aspects of the resulting release.
Throughout, $\mathbf{X}=(X_1,\ldots,X_p)$ denotes a complete record,
including the substantive response $Y$, and $\mathbf{Z}$ denotes its
remaining predictors. A lowercase bold symbol denotes an observed record.
The binary discriminator label $C$ is separate from $Y$: it indicates
whether a record belongs to the source or auxiliary artificial class.

\subsection{Synthetic Data Quality and the Privacy--Utility Trade-off}
\label{subsec:basic_concepts_in_syn_generation}
Synthetic data quality comprises several complementary
dimensions \citep{snoke2018}.
\textit{Statistical fidelity} concerns agreement between
real and synthetic marginal distributions, conditional
distributions and multivariate associations.
\textit{Predictive utility} concerns the preservation of
relationships required for a specified analytical task,
and can be assessed by evaluating models trained on real
and synthetic data against a common real test dataset.
\textit{Privacy protection} concerns limiting disclosure
about the original observations; empirical diagnostics
include exact matches, nearest-neighbour distances and
membership inference attacks \citep{stadler2022}.
Two additional properties characterize generated records.
\textit{Plausibility} describes their compatibility with
patterns learned from the real data, whereas
\textit{diversity} describes configurational variability
and the avoidance of excessive concentration.
Neither property alone ensures fidelity: plausible and
diverse records may still reproduce population frequencies
or associations inaccurately.

\label{subsec:synthetic_data_background}
Fully synthetic data replace all original records with
generated observations, whereas partially synthetic data
replace selected variables or sensitive components
\citep{rubin1993,reiter2005}.
Let $P_R$ denote the real-data distribution and $P_S$
the synthetic-data distribution.
Fidelity requires agreement between relevant features of
$P_S$ and $P_R$, while utility depends on the analytical task.
Distributional agreement does not itself imply memorization;
however, reproducing rare training configurations may expose
individual-level information.
Conversely, mechanisms that discourage such configurations
can alter relationships relevant to analysis.
Differential privacy formalizes protection by bounding
the influence of individual records on a randomized
procedure \citep{dwork2014}.
Distance-based sampling penalties instead discourage
proximity to observed records without providing the same
formal guarantee.
Their effects on disclosure risk and utility therefore
require empirical assessment.

\subsection{Discriminative Learning}
\label{subsec:discriminative_learning_background}
Discriminative learning estimates a response distribution conditional
on observed predictors. For the plausibility discriminator used here,
the response is the binary class indicator $C$, and its inputs are all
coordinates of a candidate record $\mathbf{x}$:
\begin{equation}
p_{\mathrm{real}}(\mathbf{x})
=\Pr(C=1\mid\mathbf{X}=\mathbf{x}).
\label{eq:discriminative_learning}
\end{equation}
Here $C=1$ denotes the source class and $C=0$ the auxiliary artificial
class. This task differs from predicting the substantive response $Y$
from $\mathbf{Z}$. Under a logistic formulation,
\begin{equation}
p_{\mathrm{real}}(\mathbf{x})=\sigma(f(\mathbf{x})),
\label{eq:logistic_model}
\end{equation}
where $\sigma(u)=1/(1+\exp(-u))$ and $f(\mathbf{x})$ is a real-valued score.
Gradient boosting constructs this score through an additive
ensemble of sequentially fitted learners, typically
decision trees \citep{chen2016xgboost}.
A discriminator trained to distinguish real records from
artificially constructed records can supply a plausibility
score for candidate observations.
Larger real-class scores indicate greater resemblance to
patterns identified by the fitted discriminator.
Their interpretation depends on the construction of the
artificial class and the training class proportions;
they are not absolute probabilities of plausibility.
Once fitted, the discriminator can remain fixed during
stochastic generation, providing a record-level criterion
without joint adversarial training.

\subsection{Bayesian Networks}
\label{subsec:bayesian_networks_background}
A Bayesian network represents a joint distribution through
a directed acyclic graph (DAG) \citep{scutari2010}.
The nodes are the coordinates of the complete record $\mathbf{X}$,
including $Y$, rather than only the predictors of a classification task.
Writing $q_{\mathrm{BN}}$ for its joint probability mass function gives
\begin{equation}
q_{\mathrm{BN}}(\mathbf{x})
=\prod_{j=1}^{p}q_j(x_j\mid\mathbf{x}_{\mathrm{pa}(j)}),
\label{eq:bn_factorization}
\end{equation}
where $x_{\mathrm{pa}(j)}$ denotes the configuration of
the parent nodes of $X_j$.
The graph encodes conditional independence relationships,
allowing the joint distribution to be represented through
local conditional models $q_j$. Fitting these models to the categorical
training reference produces the proposal distribution used in
Section~\ref{subsec:mh_generation}. Thus the network supplies complete
candidate records, whereas the discriminator scores their resemblance
to the source class.
For categorical variables, these models specify
probabilities over the respective variable domains.
Candidate records can be generated in topological order
by sampling each variable conditionally on its parents.
Unlike independent sampling from marginal distributions,
this mechanism incorporates fitted dependencies directly
into proposals.
Its representational complexity depends on the graph
structure and variable cardinalities, and the fitted
network need not capture every association in the data.

\subsection{Metropolis--Hastings Sampling}
\label{subsec:mh_background}
Metropolis--Hastings is a Markov chain Monte Carlo method
for sampling from a target distribution $\pi$
\citep{metropolis1953,hastings1970,chib1995}.
Given the current state $x^{(t)}$, a candidate $x^\star$
is drawn from
\begin{equation}
q(x^\star \mid x^{(t)}).
\end{equation}
Writing $x$ for the current state, the acceptance
probability is
\begin{equation}
\label{eq:mH_acceptance}
\alpha(x,x^{\star})
=
\min\left\{
1,
\frac{
\pi(x^{\star})\,q(x\mid x^{\star})
}{
\pi(x)\,q(x^{\star}\mid x)
}
\right\}.
\end{equation}
If the proposal is rejected, the current state is retained.
The ratio
\begin{equation}
\frac{q(x^{(t)}\mid x^\star)}
{q(x^\star\mid x^{(t)})}
\end{equation}
is the Hastings correction, accounting for proposal
asymmetry.
It equals one for symmetric proposals; independence
from the current state does not itself imply symmetry.
For a fixed target, the resulting transition kernel
satisfies detailed balance:
\begin{equation}
\pi(x)\,P(x \rightarrow x^\star)
=
\pi(x^\star)\,P(x^\star \rightarrow x),
\end{equation}
where $P(x \rightarrow x^\star)$ denotes the transition
probability.
Detailed balance establishes invariance of $\pi$.
On a finite state space, irreducibility and aperiodicity
ensure convergence to the invariant distribution.
Finite runs may nevertheless retain initialization
effects and dependence between successive states.
Furthermore, this argument concerns a fixed target;
modifying the target during sampling requires a separate
assessment.
Because acceptance depends on probability ratios,
the normalizing constant of $\pi$ need not be computed.

\subsection{Energy-Based Modelling}
\label{subsec:energy_background}
Energy-based models associate a configuration $x$
with an energy $H(x)$ through \citep{lecun2006}
\begin{equation}\label{eq:energy_prob}
\pi_H(x)
=
\frac{\exp\{-H(x)\}}{Z},
\end{equation}
where $Z$ is the partition function.
On a discrete state space, it is the sum of the
unnormalized probabilities over admissible configurations.
Lower-energy configurations consequently have higher
probability.
Multiple criteria can be combined as
\begin{equation}
H(x)
=
\sum_{r=1}^{R}
\lambda_r H_r(x),
\label{eq:multiobjective_energy}
\end{equation}
where $H_r(x)$ is a component penalty and
$\lambda_r \geq 0$ determines its contribution.
This linear scalarization maps potentially competing
objectives to a single quantity.
Component scales and coefficients jointly determine
their influence, allowing improvements in one component
to compensate for deterioration in another.
Finite penalties encourage properties without imposing
hard constraints; hard exclusions instead remove
configurations from the admissible state space.
The configuration $x$ may represent either a record
or an entire dataset, allowing sampling and selection
to use distinct objectives within the same formalism.

\subsection{Simulated Annealing}
\label{subsec:sa_background}
Simulated annealing minimizes an energy function through
stochastic transitions governed by a decreasing
temperature \citep{kirkpatrick1983,aarts1989}.
For a symmetric proposal from $x$ to $x^\star$,
the acceptance probability is
\begin{equation}
\alpha_{\mathrm{SA}}
=
\min\left\{
1,
\exp\left[
-\frac{H(x^\star)-H(x)}{T}
\right]
\right\},
\end{equation}
where $T>0$.
Asymmetric proposals require the corresponding
Hastings correction.
Under the symmetric rule, energy-decreasing proposals
are accepted, while energy-increasing proposals may
be accepted according to their energy difference
and the temperature.
Higher temperatures facilitate exploration; cooling
progressively favours local improvement.
Asymptotic convergence to global minima requires
appropriate conditions on the state space, proposal
mechanism and cooling schedule.
It does not imply global optimality for a finite run.
The method requires no derivatives and is therefore
applicable to discrete problems such as fixed-size
selection from a candidate pool.

\subsection{Privacy Diagnostics and Membership Inference Attacks}
\label{subsec:privacy_background}
Let
$\mathcal{D}_{\mathrm{real}}=\{\mathbf{x}_i\}_{i=1}^{n_r}$
denote the real dataset and
$\mathcal{D}_{\mathrm{syn}}=\{\tilde{\mathbf{x}}_j\}_{j=1}^{n_s}$
the synthetic dataset.
An exact match occurs when
$\tilde{\mathbf{x}}_j=\mathbf{x}_i$.
Such matches describe overlap but do not alone establish
memorization, since different individuals may share a
configuration.
Conversely, their absence does not establish privacy.
For a synthetic observation $\tilde{\mathbf{x}}$,
proximity to the real data can be measured by
the minimum normalized Hamming distance:
\begin{equation}
d_{\min}(\tilde{\mathbf{x}})
=
\min_{x_i \in \mathcal{D}_{\mathrm{real}}}
\frac{1}{p}
\sum_{k=1}^{p}
\mathbb{I}
(\tilde{x}_k \neq x_{ik}),
\end{equation}
where $p$ is the number of variables and
$\mathbb{I}(\cdot)$ is the indicator function.
This distance measures coordinate disagreement,
not the magnitude of numerical differences.
Diversity can similarly be described through
pairwise synthetic distances:
\begin{equation}
d(\tilde{x}_i,\tilde{x}_j)
=
\frac{1}{p}
\sum_{k=1}^{p}
\mathbb{I}
(\tilde{x}_{ik} \neq \tilde{x}_{jk}).
\end{equation}
Larger distances indicate greater configurational
variation, which should be interpreted jointly with
fidelity because dispersion can also reflect
distributional distortion.
Marginal fidelity can be measured by
\begin{equation}
\Delta_v
=
\frac{1}{2}
\sum_{l}
|p_{\mathrm{real}}(l)-p_{\mathrm{syn}}(l)|,
\end{equation}
where $l$ indexes the levels of variable $v$
on a common support.
This is the total variation distance (TVD); analogous
comparisons apply to bivariate and conditional
distributions.
Membership inference attacks assess whether an adversary
can distinguish records used during fitting or generation
from excluded records \citep{shokri2017,salem2019}.
Attacks may use distance thresholds or supervised
classifiers fitted to record-level features.
Attack fitting, threshold calibration and evaluation
require separate candidate partitions.
Performance is summarized through the area under
the ROC curve (AUC), true positive rate (TPR),
false positive rate (FPR), and attack advantage:
\begin{equation}
\mathrm{Advantage}
=
\mathrm{TPR}-\mathrm{FPR}.
\end{equation}
Performance near random guessing indicates limited
membership discrimination for the evaluated attacks
and candidate population.
It does not exclude disclosure under other attacks
or information assumptions.


\section{Proposed Approach}
\label{sec:proposed_approach}

The framework separates record-level generation from
dataset-level selection.
The \textit{pool-construction phase} constructs a candidate pool using
Bayesian-network proposals and energy-driven acceptance
guided by discriminative plausibility, privacy and diversity.
The \textit{target-selection phase} selects a fixed-size target by
minimizing distributional and classification-accuracy
discrepancies.
These explicit criteria define the open-box formulation
summarized in Figure~\ref{fig:pipeline}.

\subsection{Data Representation and Preliminary Learning}
\label{subsec:preprocessing_learning}

Let
\begin{equation}
\mathcal{D}_R
=
\{
\mathbf{x}_1,
\mathbf{x}_2,
\ldots,
\mathbf{x}_n
\}
\end{equation}
denote the real reference dataset of $n$ records, each containing $p$
nominal, ordinal or numerical variables. The count $p$ includes the
substantive response: each record can be written as
$\mathbf{x}_i=(\mathbf{z}_i,y_i)$, with $p-1$ predictors.
A categorical response $Y$ defines the classification
task, and $\mathbf{Z}$ denotes the remaining predictors.
All models, discretization intervals and reference
distributions are estimated from the real training data;
a separate real holdout is reserved for external evaluation.
A common preprocessing policy, including the treatment
of missing values and declared missing-value labels,
defines the reference for all components.
Numerical predictors are discretized through quantile-based
intervals, while categorical variables retain their domains.
The boundaries and interval-closure conventions are saved
for reconstruction.
The resulting categorical reference $\mathcal{D}_R^c$
and finite configuration space $\mathcal{X}$ support
proposal generation, Hamming distances and distributional
comparisons.
Preliminary learning fits models with three distinct roles.
First, the XGBoost discriminator of
Section~\ref{subsec:discriminative_learning_background} supplies a
record-level plausibility score. Second, the Bayesian network of
Section~\ref{subsec:bayesian_networks_background} supplies proposals and
the baseline probability structure of pool construction. Third, Random
Forests select informative variable pairs and provide the fixed
classification guide for target selection. Pair screening and the guide
are separate fits: the former identifies distributions to compare,
whereas the latter measures one task-specific accuracy discrepancy.
External evaluation later trains new classifiers on the released data.

A \emph{screening response} is simply a variable temporarily treated as
the outcome of a pair-selection forest. Several variables can play this
role; it does not redefine the substantive response $Y$.
For bivariate selection, each designated screening response
is predicted from the eligible predictors, excluding itself,
and its highest-ranked predictor defines a candidate pair.
Screening responses and predictor exclusions are specified
before fitting.
Reciprocal selections are retained once, producing a set
$\mathcal{B}$ of unordered pairs.
Importance rankings identify associations but do not
determine their weights in the fidelity objective.

\subsection{Discriminative Learning and Pool-Construction Penalties}
\label{subsec:flight_energy}

A logistic XGBoost discriminator contrasts the categorical
real reference with an auxiliary artificial dataset
$\mathcal{D}_F$, which supplies the negative training class.
Artificial records combine independent draws from empirical
marginals with modified copies of training records.
Modifications comprise permutations of selected columns,
resampling selected coordinates from their empirical marginals,
and replacing selected values with alternative categories.
These mechanisms disrupt dependencies to different degrees,
providing both broadly recombined and locally modified
configurations rather than a uniformly implausible class.
Exact matches with the real reference and duplicate artificial
records are removed from this discriminator-training sample.
Additional independent marginal draws replenish the sample
to its prescribed size.
This filtering concerns only $\mathcal{D}_F$; duplicate
contents remain admissible in the subsequent synthetic pool.
All artificial-class construction uses the training reference,
without accessing the real holdout.
The fitted output $\hat p_{\mathrm{real}}(\mathbf{x})$ estimates the
source-class probability defined in
Equation~\ref{eq:discriminative_learning}, using the source and auxiliary
records labelled $C=1$ and $C=0$, respectively.
For the fitted real-class score $\hat p_{\mathrm{real}}(\mathbf{x})$, the
plausibility energy is
\begin{equation}
H_{\mathrm{plaus}}(\mathbf{x})
=
1-\hat p_{\mathrm{real}}(\mathbf{x}).
\end{equation}
Since $\hat p_{\mathrm{real}}(\mathbf{x})\in[0,1]$, this penalty lies in $[0,1]$
and decreases with the real-class score.
Its interpretation depends on the artificial-class
construction, and the discriminator remains fixed
throughout pool construction.
Let $A_t$ denote the synthetic archive at iteration $t$
and $d_H(A_t,\mathbf{x})$ the minimum normalized Hamming
distance from the candidate to that archive.
For any nonempty collection $A$ of categorical records, define
\begin{equation}
d_H(A,\mathbf{x})=
\min_{\mathbf{a}\in A}\frac{1}{p}\sum_{j=1}^{p}
\mathbb{I}\{a_j\neq x_j\}.
\label{eq:hamming_collection}
\end{equation}
Thus $d_H$ counts the fraction of differing coordinates to the closest
record, including the response coordinate. Numerical variables enter
through their bins; the distance does not measure differences in years
or other original units. It is the minimum-distance construction of
Section~\ref{subsec:privacy_background} applied to an arbitrary collection.
The diversity penalty is
\begin{equation}
H_{\mathrm{div}}(\mathbf{x};A_t)
=
\exp
\left(
-
\frac{
d_H(A_t,\mathbf{x})
}{
\tau_D
}
\right),
\end{equation}
where $\tau_D>0$ controls the distance scale.
The penalty is zero for an empty archive and otherwise
discourages proximity to previously collected configurations.
Duplicate synthetic contents remain admissible.
The privacy penalty similarly uses the minimum distance
to the categorical real reference:
\begin{equation}
H_{\mathrm{priv}}(\mathbf{x})
=
\exp
\left(
-
\frac{
d_H(\mathcal{D}_R^c,\mathbf{x})
}{
\tau_R
}
\right),
\end{equation}
where $\tau_R>0$.
Exact categorical matches are additionally excluded,
defining the admissible space
\begin{equation}
\mathcal{X}_0
=
\left\{
\mathbf{x}\in\mathcal{X}:
d_H(\mathcal{D}_R^c,\mathbf{x})>0
\right\}.
\end{equation}
The combined penalty is
\begin{equation}
E(\mathbf{x};A_t)
=
\lambda_P H_{\mathrm{plaus}}(\mathbf{x})
+
\lambda_R H_{\mathrm{priv}}(\mathbf{x})
+
\lambda_D H_{\mathrm{div}}(\mathbf{x};A_t).
\label{eq:flight_penalty}
\end{equation}
The non-negative coefficients $\lambda_P$, $\lambda_R$
and $\lambda_D$ remain fixed during a run.
Their relative values determine the balance among
components, while their overall scale affects acceptance
selectivity; they need not sum to one.
The distance scales $\tau_R$ and $\tau_D$ control
how rapidly the respective penalties decrease.

\subsection{Bayesian-Network Proposals and Energy-Driven Sampling}
\label{subsec:mh_generation}

This step combines the network factorization of
Equation~\ref{eq:bn_factorization}, the energy representation of
Section~\ref{subsec:energy_background}, and the acceptance rule of
Section~\ref{subsec:mh_background}. 
The pool construction uses independence
Metropolis--Hastings sampling with Bayesian-network proposals
and an archive-dependent, energy-reweighted target distribution. 
Recall that $\mathcal{D}_R^c$ is the discretized training reference,
$\mathcal{X}_0$ excludes its exact categorical matches, $A_t$ is the
archive of previously collected synthetic records, and
$E(\mathbf{x};A_t)$ combines plausibility, reference-proximity and
archive-proximity penalties.
Let $q_{\mathrm{BN}}(\mathbf{x})$ denote the joint
distribution of the Bayesian network fitted to
$\mathcal{D}_R^c$.
A candidate $\mathbf{x}^{\star}$ is drawn independently
of the current state $\mathbf{x}$:
\begin{equation}
q(\mathbf{x}^{\star}\mid\mathbf{x})
=
q_{\mathrm{BN}}(\mathbf{x}^{\star}).
\end{equation}
The same distribution contributes its negative
log-probability to the pool-construction energy. Its role is to favour
configurations supported by the fitted dependencies, while the added
penalties reduce preference for configurations that score poorly on the
three explicit criteria:
\begin{equation}
H_{\mathrm{pool}}(\mathbf{x};A_t)
=
-\log q_{\mathrm{BN}}(\mathbf{x})
+
E(\mathbf{x};A_t),
\qquad
\mathbf{x}\in\mathcal{X}_0.
\label{eq:flight_energy}
\end{equation}
For a fixed archive, the corresponding target is
\begin{equation}
\pi_{A_t}(\mathbf{x})
=
\frac{
q_{\mathrm{BN}}(\mathbf{x})
\exp\{-E(\mathbf{x};A_t)\}
}{
Z(A_t)
},
\qquad
\mathbf{x}\in\mathcal{X}_0,
\label{eq:flight_target}
\end{equation}
where $Z(A_t)$ is the normalizing constant.
Intuitively, this target reweights the network distribution: the factor
$\exp\{-E(\mathbf{x};A_t)\}$ reduces the relative probability of penalized
records. The normalizer makes the resulting probabilities sum to one
over the admissible support. The target depends on $A_t$ because diversity
is assessed relative to records already collected.
Configurations outside $\mathcal{X}_0$ have zero
target probability.
The network therefore supplies both proposals and
the reference probability structure modified by
the explicit penalties.
For current and proposed states on the admissible
positive-probability support, the generally asymmetric
independence proposal gives
\begin{equation}
\alpha_{A_t}(\mathbf{x},\mathbf{x}^{\star})
=
\min\left\{
1,
\frac{
q_{\mathrm{BN}}(\mathbf{x}^{\star})
\exp\{-E(\mathbf{x}^{\star};A_t)\}
}{
q_{\mathrm{BN}}(\mathbf{x})
\exp\{-E(\mathbf{x};A_t)\}
}
\frac{
q_{\mathrm{BN}}(\mathbf{x})
}{
q_{\mathrm{BN}}(\mathbf{x}^{\star})
}
\right\}.
\label{eq:flight_hastings}
\end{equation}
The Bayesian-network factors cancel with the Hastings
correction, yielding
\begin{equation}
\alpha_{A_t}(\mathbf{x},\mathbf{x}^{\star})
=
\min\left\{
1,
\exp\left[
E(\mathbf{x};A_t)
-
E(\mathbf{x}^{\star};A_t)
\right]
\right\}.
\label{eq:energy_acceptance}
\end{equation}
Candidates outside the admissible space are rejected.
Cancellation requires the same network distribution
in the proposal and target, with coefficient one
on its negative log-probability; it does not require
proposal symmetry.
Although its probability ratio need not be evaluated,
the network still determines proposal frequencies
and remains part of Equation~\ref{eq:flight_target}.
Current and proposed records are evaluated against
the same archive $A_t$.
Plausibility and distance to the real reference can
be cached for each configuration, whereas diversity
must reflect the archive used in the comparison.
After burn-in, states are collected at a specified
thinning interval until the required pool size is reached.
Rejected transitions retain the current state, which
may consequently be collected more than once.
No post-generation deduplication is applied.
Archive updates alter the diversity penalty and target:
detailed balance holds with the archive fixed, rather
than establishing a single stationary distribution
for the complete adaptive procedure.
Parallel chains maintain separate archives and
random-number streams.
Their collected records form the pool, with diversity
controlled within each chain during generation.

\subsection{Target Selection: Fidelity and Predictive Utility}
\label{subsec:landing_phase}
Target selection addresses distributional discrepancies that
record-level pool construction criteria do not directly control.
For a pool of $M$ indexed records and target size $m<M$,
the selected index set satisfies
\begin{equation}
I\subseteq\{1,\ldots,M\},
\qquad
|I|=m.
\end{equation}
Selection is without replacement at the index level,
although different indices may contain identical records.
The categorical target $\mathcal{D}_S^c(I)$ is initialized
by uniformly sampling $m$ pool indices.
Each proposal uniformly selects one included index
for removal and one excluded index for inclusion,
without modifying record contents.
The objective is
\begin{equation}
H_L(\mathcal{D}_S^c)
=
H_F(\mathcal{D}_S^c)
+
H_U(\mathcal{D}_S^c),
\label{eq:landing_energy}
\end{equation}
with unit coefficients for fidelity and utility.
Selection remains restricted to the pool-generated pool;
the properties of the selected release are assessed
subsequently.

\subsubsection{Distributional Fidelity}
\label{subsec:landing_fidelity}
Fidelity is evaluated on the categorical representation,
including numerical-predictor bins.
The following discrepancies apply the TVD definition from
Section~\ref{subsec:privacy_background} to three objects: single-variable
frequencies, joint frequencies of selected pairs, and within-stratum
frequencies. The set $\mathcal{B}$ is formed by the RF pair screening in
Section~\ref{subsec:preprocessing_learning}: one top-ranked predictor is
selected for each designated screening response and reciprocal pairs are
merged. All probabilities below are empirical frequencies on the same
categorical support.
For variable $j$, the marginal total variation distance is
\begin{equation}
d_j(\mathcal{D}_S^c)
=
\frac{1}{2}
\sum_{k=1}^{K_j}
\left|
p^{real}_{jk}
-
p^{syn}_{jk}(\mathcal{D}_S^c)
\right|,
\label{eq:dj}
\end{equation}
where $K_j$ is the number of reference categories.
For $(j,k)\in\mathcal{B}$, the bivariate discrepancy is
\begin{equation}
d_{jk}(\mathcal{D}_S^c)
=
\frac{1}{2}
\sum_{a=1}^{K_j}
\sum_{c=1}^{K_k}
\left|
p^{real}_{jk}(a,c)
-
p^{syn}_{jk}(a,c;\mathcal{D}_S^c)
\right|.
\label{eq:landing_pair}
\end{equation}
Let $b$ denote the variable with the largest cardinality
in the categorical reference.
This is an operational stratification rule: it chooses the finest
single-variable partition available in that representation. In this
application it selects municipality, so the conditional terms assess
whether distributions are preserved within geographic groups. Cardinality
alone does not establish scientific relevance; in other applications the
conditioning variable should be specified from the analytical purpose,
and sparse strata can make these discrepancies difficult to estimate.
For each $j\neq b$, conditional fidelity is measured by
\begin{equation}
d_{j\mid b}(\mathcal{D}_S^c)
=
\sum_{g=1}^{K_b}
p_b^{real}(g)
\frac{1}{2}
\sum_{a=1}^{K_j}
\left|
p^{real}(x_j=a\mid x_b=g)
-
p^{syn}(x_j=a\mid x_b=g;\mathcal{D}_S^c)
\right|.
\label{eq:landing_conditional}
\end{equation}
Reference stratum proportions remain fixed.
If a reference stratum is absent from the target,
its inner conditional discrepancy is assigned one.
All comparisons use the same reference support.
The fidelity energy is the arithmetic mean
\begin{equation}
H_F(\mathcal{D}_S^c)
=
\frac{
\displaystyle\sum_{j=1}^{p}d_j(\mathcal{D}_S^c)
+
\displaystyle\sum_{(j,k)\in\mathcal{B}}
d_{jk}(\mathcal{D}_S^c)
+
\displaystyle\sum_{j\neq b}
d_{j\mid b}(\mathcal{D}_S^c)
}{
p+|\mathcal{B}|+(p-1)
}.
\label{eq:hmean}
\end{equation}
Each marginal, bivariate and conditional component
receives equal weight and lies in $[0,1]$;
hence $H_F\in[0,1]$.
Equal component weights provide a transparent default because all terms
share the same range, without introducing additional priorities from
importance scores. They do not give equal total weight to the three
families: their contributions depend on their numbers of terms,
$p$, $|\mathcal{B}|$ and $p-1$. Nor does equal range imply equal sampling
variability. These are declared design choices, rather than an optimal
weighting result; application-specific priorities may motivate other
weights.

\subsubsection{Predictive Utility}
\label{subsec:classification_utility}
A Random Forest classifier $\hat f$ is fitted to the
real reference using categorical predictors and remains
fixed throughout target selection.
For reference predictors $\mathbf{z}_i^c$ and response
$y_i$, its accuracy is
\begin{equation}
Acc_{real}
=
\frac{1}{n}
\sum_{i=1}^{n}
\mathbb{I}
\left\{
\hat f(\mathbf{z}_i^c)=y_i
\right\},
\end{equation}
and its accuracy on the current target is
\begin{equation}
Acc_{syn}(\mathcal{D}_S^c)
=
\frac{1}{m}
\sum_{i=1}^{m}
\mathbb{I}
\left\{
\hat f(\tilde{\mathbf{z}}_i^c)
=
\tilde y_i
\right\}.
\end{equation}
The utility energy is
\begin{equation}
H_U(\mathcal{D}_S^c)
=
\left|
Acc_{syn}(\mathcal{D}_S^c)
-
Acc_{real}
\right|.
\label{eq:hrf}
\end{equation}
This discrepancy lies in $[0,1]$.
The reference accuracy is computed on the guide's
training records and serves as an internal optimization
benchmark, distinct from external holdout performance.
Pool predictions are computed once, allowing target
accuracy to be updated without refitting.

\subsubsection{Annealing and Reconstruction}
\label{subsec:landing_annealing}
The simulated-annealing rule of Section~\ref{subsec:sa_background}
is now applied to whole selected datasets, with energy $H_L$, rather
than to individual candidate records.
The uniform index-swap proposal has the same probability
as its reverse.
A proposed target $S^\star$ is therefore accepted
from the current target $S$ with probability
\begin{equation}
\alpha_L(S,S^\star)
=
\min\left\{
1,
\exp\left[
-\frac{
H_L(S^\star)-H_L(S)
}{
T_t
}
\right]
\right\},
\label{eq:landing_acceptance}
\end{equation}
where $T_t>0$ is the temperature at iteration $t$.
Preliminary swaps calibrate the initial temperature,
followed by geometric cooling.
The iteration budget and cooling parameters are fixed
for each run, which returns the final chain state.
Attainable discrepancies depend on pool support
and the finite optimization budget.
After target selection, numerical predictors are sampled uniformly
within their selected intervals; integer-valued predictors
are sampled uniformly over the admissible integers.
Saved boundaries and interval-closure conventions ensure
that reapplying the discretization recovers the selected
categorical record, preserving the fidelity assessed
during target selection.
Categorical coding remains unchanged, and reconstructed
records are not deduplicated.

\subsection{Diagnostic Assessment}
\label{subsec:diagnostics}

Assessment uses the reconstructed release.
Fidelity diagnostics examine marginal, selected bivariate
and conditional distributions, verify bin preservation
and assess numerical agreement on the reconstructed scale.
External utility compares identical classification-model
specifications fitted to real and synthetic training data
and evaluated on a shared real holdout.
These models are separate from the target selection guide.
Repeated classifier training runs with different random
seeds and fixed hyperparameters assess training variability
on the same synthetic dataset.
Exact matches and nearest-neighbour distances describe
proximity to real records, while duplicate counts and
within-synthetic distances characterize concentration
and diversity.
Membership inference attacks use separate candidate
partitions for fitting, calibration and testing.
Separately, repeated target selection and reconstruction
runs from a fixed pool assess variation across different
synthetic datasets.
Repeated attack partitions assess membership-inference
variability for each release.
Interpretation accounts for dependence induced by the
shared pool, common holdout and overlapping attack
candidates; privacy findings remain specific to the
evaluated attacks.
\begin{figure}[htbp]
\centering
\resizebox{\textwidth}{!}{%
\begin{tikzpicture}[
font=\small,
box/.style={
rectangle,
draw=black,
rounded corners=2pt,
align=center,
text width=3.2cm,
minimum height=1.65cm,
inner sep=5pt
},
arrow/.style={
->,
line width=0.75pt
}
]
\node[box] (real) at (0,0)
{\textbf{Real training data}\\
Categorical representation\\
Saved numerical intervals};
\node[box] (learn) at (4.2,0)
{\textbf{Preliminary learning}\\
XGB plausibility discriminator\\
BN structure and RF guides};
\node[box] (poolstep) at (8.4,0)
{\textbf{Pool construction phase}\\
BN proposal and target energy\\
MH with Hastings cancellation};
\node[box] (pool) at (12.6,0)
{\textbf{Candidate pool}\\
Plausibility, privacy, diversity\\
Duplicate contents retained};
\node[box] (selectstep) at (12.6,-3.1)
{\textbf{Target selection phase}\\
Simulated annealing\\
Fidelity + accuracy discrepancy};
\node[box] (target) at (8.4,-3.1)
{\textbf{Selected target}\\
Fixed-size categorical subset\\
Duplicate contents retained};
\node[box] (recon) at (4.2,-3.1)
{\textbf{Reconstruction}\\
Numerical predictors\\
Original bins preserved};
\node[box] (diag) at (0,-3.1)
{\textbf{Diagnostic assessment}\\
Fidelity and predictive utility\\
Privacy, diversity and stability};
\draw[arrow] (real) -- (learn);
\draw[arrow] (learn) -- (poolstep);
\draw[arrow] (poolstep) -- (pool);
\draw[arrow] (pool) -- (selectstep);
\draw[arrow] (selectstep) -- (target);
\draw[arrow] (target) -- (recon);
\draw[arrow] (recon) -- (diag);
\end{tikzpicture}%
}
\caption{Generation, selection and evaluation workflow.
Pool construction combines Bayesian-network proposals with
plausibility, privacy and diversity penalties.
Target selection optimizes fidelity and accuracy agreement;
numerical reconstruction preserves the selected bins
before external assessment.}
\label{fig:pipeline}
\end{figure}


\section{Case Study}
\label{sec:case_study}

The empirical application uses an artificially constructed
dataset of $10000$ individual-level records and $p=8$
variables, representing a simulated hospital-triage setting.
The records do not correspond to observations collected
from actual patients.
The four-category response \texttt{diagnosis} is generated
deterministically from age, physical activity and genetic
predisposition.
After min--max normalization, two scores combine age
(weight $0.2$) with one minus normalized physical activity
and with normalized genetic predisposition, respectively
(weight $0.8$).
Each score is dichotomized at its median in the original
$10000$-record dataset, assigning $1$ to values at or above
the median and $0$ otherwise.
The response is then defined as $2b_1+b_2+1$, where $b_1$
and $b_2$ are the resulting binary indicators.
The four combinations therefore define classes $1,2,3,4$.
The supplied generation notebook applies no subsequent
random relabelling of the response.
The intermediate scores and binary indicators are excluded
from the eight variables used in this study.

This controlled setting supports assessment of distributional
and predictive preservation and training-membership disclosure.
Throughout the study, ``real'' denotes these source data,
as distinguished from the synthetic records produced by
the proposed method.
Before preprocessing and model fitting, the dataset is
partitioned into a training reference $\mathcal{D}_R$,
with $|\mathcal{D}_R|=9000$, and a holdout
$\mathcal{D}_{\mathrm{test}}$ of $1000$ records.
This partition remains fixed throughout the evaluation
of the proposed method.
The training reference supplies discretization boundaries,
Bayesian-network estimation, XGBoost plausibility modelling,
RF-based pair selection, the fixed classification guide,
reference distributions and privacy distances.
The holdout is reserved for external evaluation.
The variable \texttt{municipality\_residence} contains
$145$ categories in the preprocessed training reference.
Rare municipality levels create sparse configurations
that complicate empirical distribution matching and
record-proximity assessment.
Table~\ref{tab:dataset_characteristics} summarizes
the dataset and experimental partition.
\begin{table}[htbp]
\caption{Main characteristics of the source dataset and
experimental partition. Predictor counts exclude the response.}
\label{tab:dataset_characteristics}
\centering
\begin{tabular}{lr}
\toprule
Characteristic & Value \\
\midrule
Total source records & 10000 \\
Training reference records & 9000 \\
Holdout records & 1000 \\
Records per synthetic target & 9000 \\
Variables, including the response & 8 \\
Municipality categories in the training reference & 145 \\
Nominal predictors & 3 \\
Ordinal predictors & 2 \\
Binary predictor & 1 \\
Numerical predictors & 1 \\
Response variable & Diagnosis \\
Requested age discretization bins & 20 \\
\bottomrule
\end{tabular}
\end{table}
Table~\ref{tab:variables_case_study} reports the variables,
their types and interpretations.
Nominal variables are unordered, whereas ordinal
predictors retain their ordered interpretation.
\begin{table}[htbp]
\caption{Variables used in the case study.}
\label{tab:variables_case_study}
\centering
\begin{tabular}{lll}
\toprule
Variable & Type & Description \\
\midrule
\texttt{municipality\_residence}
& Nominal & Municipality of residence \\
\texttt{gender}
& Nominal & Gender of the individual \\
\texttt{civil\_status}
& Nominal & Marital or civil status \\
\texttt{occupation}
& Binary & Occupational condition (codes 0 and 1) \\
\texttt{physical\_activity}
& Ordinal & Level of physical activity \\
\texttt{genetic\_predisposition}
& Ordinal & Genetic predisposition category \\
\texttt{age}
& Numerical & Age of the individual \\
\texttt{diagnosis}
& Categorical response & Class of the health condition \\
\bottomrule
\end{tabular}
\end{table}
The response \texttt{diagnosis} defines the multiclass
classification task used in target selection and external
validation.
It has four classes, coded $1,2,3,4$ in the saved reference.
The occupation variable has two observed codes, $0$ and $1$;
it is reported as binary rather than assigning an unsupported
substantive ordering to occupational categories.
Its ordered-factor storage does not by itself establish
such an interpretation. 
The constructed relationships between \texttt{diagnosis}
and \texttt{genetic\_predisposition},
\texttt{physical\_activity} and \texttt{age}
allow predictive preservation to be assessed beyond
marginal agreement.
Diagnosis is included as a variable in the joint synthetic
generation; its defining rule is not reapplied to assign
labels to the generated records.
External classification therefore evaluates whether the
synthetic data preserve usable diagnosis--covariate
relationships. 
The three external classification models are each
trained on the 9000 reference records and, separately,
ten times on the same 9000-record synthetic dataset,
using different random seeds and fixed hyperparameters. 
All fitted classifiers are evaluated on the same $1000$
source holdout records, which are excluded from synthetic
generation and classifier training.
Comparisons with the reference-trained classifiers thus
assess preservation of predictive utility across three
models, as reported in the results tables.
This evaluation concerns predictive performance rather
than exact record-level compliance with the diagnosis
generation rule or recovery of the full conditional
distribution $\Pr(Y\mid\mathbf{Z})$.

The sole numerical predictor, \texttt{age}, is recorded
in integer years and discretized using training-reference
quantiles, with $20$ requested bins and consistent handling
of coincident boundaries.
After target selection, integer ages are reconstructed within
their selected intervals so that reapplying the saved
discretization recovers the original bins.
External evaluation uses these reconstructed values.
Each synthetic target contains $9000$ records selected
from the candidate pool, with duplicate contents retained.
The external utility assessment is distinct from the fixed
RF criterion: the guide's reference accuracy is measured
on its training records and serves as an optimization benchmark. 
For membership inference, training-reference records
supply member candidates and holdout records supply
nonmember candidates.
The attacks therefore assess disclosure of training inclusion
within the controlled experiment, rather than disclosure
about actual patients.
The experimental settings were chosen operationally
and held fixed for the main experiment, without a claim
of optimality.
Pool construction used $\lambda_P=\lambda_R=1$, $\lambda_D=0.70$
and $\tau_R=\tau_D=0.25$, with a burn-in of $1000$
transitions per chain and collection every fifth
subsequent transition.
The Bayesian-network structure was learned by hill climbing
with the BIC score, $50$ random restarts and $10$
perturbations, subject to a directional blacklist.
The blacklist prevented arcs from later to earlier groups
in the ordering: municipality, gender and age;
civil status and occupation; genetic predisposition
and physical activity; diagnosis.
These constraints specified a modelling order rather
than an empirically established causal structure.
Conditional probabilities were estimated with Bayesian
smoothing and an imaginary sample size of $10$.
The discriminator's artificial training class contained
$15000$ records.
Its initial construction allocated $40\%$ to independent
empirical-marginal draws, $30\%$ to sampled reference records
with four columns permuted, $20\%$ to sampled reference
records with three coordinates resampled from their
empirical marginals, and $10\%$ to sampled reference records
with two to four coordinates replaced by alternative
categories.
Exact reference matches and duplicate artificial records
were removed; additional independent marginal draws
restored the prescribed sample size.
The percentages therefore describe the initial construction,
not the final composition after filtering and replenishment.
The logistic discriminator used a learning rate of $0.03$,
maximum tree depth $10$, minimum child weight $4$,
row and column subsampling fractions of $0.85$,
split penalty $0.5$ and $L_2$ regularization $1.5$.
Training allowed up to $900$ boosting rounds, with
an internal $80/20$ training-validation split and
early stopping after $50$ rounds without improvement
in validation log loss.
RF screening treated civil status, occupation, genetic
predisposition, physical activity and diagnosis
as separate responses, selecting the highest-ranked
predictor for each.
Each forest used $100$ trees, impurity importance,
minimum node size $10$ and half-sample draws without
replacement; the number of candidate predictors per
split was the integer part of the square root of
the available predictor count.
Municipality was explicitly excluded from this screening
but retained in the marginal and conditional fidelity
components.
The fixed RF classification guide used $500$ trees
and minimum node size $5$.
All fitting and internal validation used only the
training reference and, for the discriminator,
its derived artificial records.
Computations were performed on a laptop equipped with an
Intel Core i5-1145G7 processor at 2.60 GHz, four physical
cores, eight logical threads and 16 GB of RAM, running
Windows 11 Pro.
Pool construction was executed through six parallel chains using
a foreach-based backend \citep{kane2013,doparallel2022}.
Each chain maintained its own synthetic archive and
random-number stream, while the fitted models and
real reference remained fixed.


\section{Results}
\label{sec:results}

Results concern a principal $9000$-record release and ten
additional targets obtained through separate target selection and
reconstruction runs from the same candidate pool.
The principal release is not included in the ten-target
summaries.
It illustrates target selection optimization and detailed diagnostics;
the additional targets support stability assessment.
External utility is evaluated through repeated classifier
training on one fixed synthetic dataset.
All release-level evaluations follow numerical reconstruction
within the original age intervals.

\subsection{Generation Protocol and Candidate-Pool Results}
\label{subsec:flight_results}
Six parallel chains produce a pool of $110000$ records,
retaining duplicate configurations.
Current and proposed states are evaluated against the same
archive, as specified in Section~\ref{subsec:mh_generation}.
Table~\ref{tab:flight_summary} reports $509679$ accepted
transitions out of $556000$, corresponding to approximately
$91.67\%$.
The longest chain takes approximately $30$ minutes,
excluding preliminary model fitting.
This acceptance rate does not establish convergence
or effective sample independence.
\begin{table}[htbp]
\caption{Pool-construction execution summary for the evaluated synthetic pool.}
\label{tab:flight_summary}
\centering
\begin{tabular}{lr}
\toprule
Characteristic & Value \\
\midrule
Real reference records & 9000 \\
Parallel chains & 6 \\
Retained pool records & 110000 \\
Total transitions & 556000 \\
Accepted transitions & 509679 \\
Aggregate acceptance rate & 0.9167 \\
Longest chain duration, minutes & 29.98 \\
Mean marginal TVD (total variation distance) & 0.01280 \\
Maximum marginal TVD & 0.02836 \\
\bottomrule
\end{tabular}
\end{table}
The largest pool marginal TVD is $0.02836$ for
\texttt{municipality\_residence}, with its $145$ categories.
The remaining values range from $0.00134$ for
\texttt{gender} to $0.01669$ for \texttt{occupation}.
Marginal agreement is therefore close, although it does
not establish equivalent preservation of bivariate or
conditional distributions.

\subsection{Target selection Optimization}
\label{subsec:landing_results}
The principal target selection starts from $9000$ randomly selected
pool indices and performs $10^6$ single-record proposals,
minimizing mean fidelity discrepancy plus the absolute
guide accuracy gap with unit coefficients.
The fidelity objective contains $20$ components:
eight marginals, five bivariate distributions and seven
distributions conditional on
\texttt{municipality\_residence}.
The selected pairs link \texttt{age} with
\texttt{civil\_status}, \texttt{occupation} and
\texttt{genetic\_predisposition}, and \texttt{diagnosis}
with \texttt{physical\_activity} and
\texttt{genetic\_predisposition}.
All are identified from the training reference before
generation.
Both the total objective and mean fidelity discrepancy
decrease (Table~\ref{tab:landing_summary}).
Final guide accuracy matches the real-reference value
of $0.75044$, making the utility penalty zero.
This concerns the internal optimization benchmark,
not holdout predictive performance.
\begin{table}[htbp]
\caption{Initial and final values for the principal target selection run. Guide accuracy is evaluated using the fixed random forest employed in the optimization objective.}
\label{tab:landing_summary}
\centering
\begin{tabular}{lrr}
\toprule
Quantity & Initial selection & Final target \\
\midrule
Records & 9000 & 9000 \\
Total objective & 0.42145 & 0.02773 \\
Mean fidelity discrepancy & 0.06178 & 0.02773 \\
Guide accuracy & 0.39078 & 0.75044 \\
Absolute guide accuracy gap & 0.35967 & 0.00000 \\
Categorical duplicate records & 50 & 148 \\
\bottomrule
\end{tabular}
\end{table}
Categorical duplicates increase from $50$ to $148$.
Target selection can therefore increase concentration despite
pool construction diversity control, since its objective contains
only fidelity and guide accuracy.

\subsection{Distributional Fidelity and Numerical Reconstruction}
\label{subsec:results_fidelity}
Marginal and conditional discrepancies decrease
substantially (Table~\ref{tab:fidelity_components_summary}).
The largest final marginal discrepancy, $0.00200$,
concerns municipality; the largest conditional discrepancy,
$0.15779$, concerns age given municipality.
\begin{table}[htbp]
\caption{Fidelity discrepancies before and after target selection. Conditional discrepancies use the real-reference municipality weights. Lower values indicate closer agreement.}
\label{tab:fidelity_components_summary}
\centering
\begin{tabular}{lrrrr}
\toprule
Component family & Initial mean & Final mean & Initial maximum & Final maximum \\
\midrule
Marginal & 0.01908 & 0.00039 & 0.05633 & 0.00200 \\
Bivariate & 0.04189 & 0.03836 & 0.06678 & 0.09556 \\
Conditional & 0.12479 & 0.05140 & 0.26461 & 0.15779 \\
\bottomrule
\end{tabular}
\end{table}
Bivariate improvement is less uniform: the mean decreases
moderately, while the maximum increases.
Discrepancies for \texttt{diagnosis} with
\texttt{physical\_activity} and
\texttt{genetic\_predisposition} rise from $0.02889$
and $0.03100$ to $0.09556$ and $0.07956$, respectively.
Aggregate improvement thus permits deterioration in
individual components.
Reapplying the saved discretization after age reconstruction
produces no bin changes among the $9000$ records,
preserving all categorical fidelity components.
On the numerical scale, age has a Wasserstein distance
of $0.03778$ years, a range-normalized distance of $0.00080$
and a Kolmogorov--Smirnov distance of $0.00389$.
The real and synthetic $5$th, $25$th, $50$th, $75$th
and $95$th percentiles coincide at $21$, $33$, $46$,
$55$ and $63$ years.
Close numerical and marginal agreement therefore coexist
with residual conditional and bivariate discrepancies.

\subsection{External Predictive Utility}
\label{subsec:external_utility_results}

External utility is evaluated using extremely randomized
trees (XT), bagged classification and regression trees
(bagged CART), and XGBoost (XGB).
Each classifier is trained on the $9000$ real reference
records to obtain the real-data baseline and separately
trained ten times on the same $9000$-record synthetic
dataset using different random seeds.
All classifiers are tested on the common
$1000$-record real holdout.
Hyperparameters remain fixed across the ten runs.
The tree ensembles contain $500$ trees, and XGB uses
$300$ boosting rounds.
Predictor encoding is fitted on each training dataset
and applied to the holdout.
Synthetic means and standard deviations therefore
describe variability due to classifier randomness,
conditional on the same synthetic training dataset,
hyperparameters and holdout.
This experiment is distinct from repeating the target
selection phase to assess variation across different
synthetic datasets.
Mean synthetic-data accuracy exceeds the corresponding
baseline by $2.46$ percentage points for XT, $1.05$ for
bagged CART and $0.50$ for XGB
(Table~\ref{tab:external_classification}).
Balanced accuracy and macro-F1 follow the same direction.

\begin{table}[htbp]
\caption{External classification performance on the common real holdout. Synthetic results are means and standard deviations across ten training runs per classifier, using different seeds on the same synthetic dataset with fixed hyperparameters. Higher values are preferable.}
\label{tab:external_classification}
\centering
\begin{tabular}{llrrr}
\toprule
Model & Metric & Real & Synthetic mean & Synthetic SD \\
\midrule
XT & Accuracy & 0.4380 & 0.4626 & 0.0034 \\
XT & Balanced accuracy & 0.4366 & 0.4615 & 0.0034 \\
XT & Macro-F1 & 0.4362 & 0.4614 & 0.0035 \\
\midrule
Bagged CART & Accuracy & 0.4670 & 0.4775 & 0.0028 \\
Bagged CART & Balanced accuracy & 0.4659 & 0.4768 & 0.0028 \\
Bagged CART & Macro-F1 & 0.4657 & 0.4766 & 0.0027 \\
\midrule
XGB & Accuracy & 0.4750 & 0.4800 & 0.0054 \\
XGB & Balanced accuracy & 0.4744 & 0.4793 & 0.0055 \\
XGB & Macro-F1 & 0.4736 & 0.4790 & 0.0054 \\
\bottomrule
\end{tabular}
\end{table}
Conversely, log-loss and multiclass Brier scores increase
for all synthetic-trained classifiers
(Table~\ref{tab:external_probability_scores}).
The Brier score sums squared probability errors over
response classes and averages them across test records.
Retained classification performance therefore accompanies
weaker predictive probability quality.
\begin{table}[htbp]
\caption{External probability-based scores on the common real holdout. Synthetic results are means and standard deviations across ten training runs per classifier, using different seeds on the same synthetic dataset with fixed hyperparameters. Lower values are preferable.}
\label{tab:external_probability_scores}
\centering
\begin{tabular}{llrrr}
\toprule
Model & Metric & Real & Synthetic mean & Synthetic SD \\
\midrule
XT & Log-loss & 1.3282 & 1.4503 & 0.0087 \\
XT & Brier score & 0.7104 & 0.7547 & 0.0035 \\
\midrule
Bagged CART & Log-loss & 1.3240 & 1.4948 & 0.0166 \\
Bagged CART & Brier score & 0.7068 & 0.7653 & 0.0041 \\
\midrule
XGB & Log-loss & 1.2849 & 1.3369 & 0.0043 \\
XGB & Brier score & 0.6897 & 0.7150 & 0.0022 \\
\bottomrule
\end{tabular}
\end{table}
The accuracy improvements are descriptive findings for
this partition and learning protocol, not evidence
of general superiority over real-data training.

\subsection{Privacy Diagnostics and Membership Inference}
\label{subsec:privacy_results}
The reconstructed principal release contains no exact
matches with the $9000$ training records.
Its $148$ categorical duplicates become $65$ released
duplicates after age reconstruction.
Nine released records match holdout records excluded
from generation; these coincidences do not demonstrate
memorization.
Absence of training matches alone, however, does not
establish protection against membership inference.
The attack evaluation uses $993$ unique member profiles
and $993$ unique nonmember profiles, excluding profiles
shared between training and holdout.
Each class is partitioned into $397$ attack-training,
$297$ calibration and $299$ test profiles.
Attacks comprise proximity and reverse-proximity scores,
nearest-neighbour density, kernel scores, exact matching
and a random-forest classifier.
A permuted-label classifier provides a negative control,
and a copy-release positive control achieves AUC $1$.
Attack selection and thresholds use calibration data;
performance is measured on separate test profiles.
For the principal release, the calibration-selected attack
yields mean test AUC $0.4984$ over twenty splits
(SD $0.0290$), balanced accuracy $0.5039$ and membership
advantage $0.0079$.
The reverse mixed-distance attack attains AUC $0.5483$
in one split but averages $0.4915$ across the twenty
partitions.
Its initial discrimination is therefore not sustained
across partitions.
These correlated repetitions assess sensitivity to attack
partitioning on a fixed release and candidate set.
Table~\ref{tab:mia_target_summary} averages the twenty-split
summaries across the ten additional targets.
Mean AUC remains approximately $0.5005$, with similar
test TPR and FPR under each threshold rule.
Nominal FPR settings refer to calibration and do not
impose exact rates on test partitions.
\begin{table}[htbp]
\caption{Membership inference results averaged across ten targets and their repeated attack splits. Attack selection and thresholds use calibration data; all reported performance measures use attack test data.}
\label{tab:mia_target_summary}
\centering
\begin{tabular}{lrrrrr}
\toprule
Threshold rule & AUC & Balanced accuracy & TPR & FPR & Advantage \\
\midrule
Youden criterion & 0.5005 & 0.5010 & 0.4512 & 0.4493 & 0.0019 \\
Nominal FPR $1\%$ & 0.5005 & 0.5004 & 0.0086 & 0.0078 & 0.0009 \\
Nominal FPR $5\%$ & 0.5005 & 0.5033 & 0.0523 & 0.0457 & 0.0066 \\
\bottomrule
\end{tabular}
\end{table}
The evaluated attacks reveal no persistent membership
advantage.
This finding concerns the releases, candidate population
and attack suite considered; it does not constitute
a formal privacy guarantee.

\subsection{Stability Across Synthetic Targets}
\label{subsec:landing_stability}
The ten additional target selection and reconstruction runs use
different target-generation seeds, retaining the same
fitted models and candidate pool.
Each selects $9000$ records through $10^6$ proposals.
Table~\ref{tab:landing_stability} shows a narrow fidelity
range of $0.02742$--$0.02869$, zero guide accuracy gaps
and no exact training matches.
Released duplicates range from $57$ to $81$, or
approximately $0.63\%$--$0.90\%$ of each target.
\begin{table}[htbp]
\caption{Stability of ten synthetic targets generated from one fixed pool. MIA AUC values are first averaged across attack splits within each target. Hamming distances are normalized by the number of variables.}
\label{tab:landing_stability}
\centering
\begin{tabular}{lrrrr}
\toprule
Quantity & Mean & SD & Minimum & Maximum \\
\midrule
Mean fidelity discrepancy & 0.02812 & 0.00044 & 0.02742 & 0.02869 \\
Maximum fidelity component & 0.15869 & 0.00146 & 0.15618 & 0.16078 \\
Absolute guide accuracy gap & 0 & 0 & 0 & 0 \\
Categorical duplicate records & 149.5 & 9.02 & 137 & 165 \\
Released duplicate records & 66.3 & 7.27 & 57 & 81 \\
Exact training matches & 0 & 0 & 0 & 0 \\
Mean nearest-real Hamming distance & 0.19394 & 0.00040 & 0.19331 & 0.19451 \\
Sampled mean within-target distance & 0.72491 & 0.00072 & 0.72376 & 0.72598 \\
Selected-attack mean MIA AUC & 0.50046 & 0.00819 & 0.49273 & 0.51823 \\
\bottomrule
\end{tabular}
\end{table}
Nearest-real and within-target distances also vary little.
The standard deviations in Table~\ref{tab:landing_stability}
describe variation across synthetic targets, conditional
on the shared reference, fitted models and candidate pool.
They complement the standard deviations in
Tables~\ref{tab:external_classification}
and~\ref{tab:external_probability_scores}, which describe
classifier training variability on one fixed synthetic dataset.
The target stability experiment does not quantify variability
from refitting the generator, repeating pool construction
or sampling new populations.

\subsection{Comparison with a Diffusion-Based Generator}
\label{subsec:tabdiff_comparison}

The available TabDiff release and the proposed framework use the same
training reference of $9000$ records and the same
held-out set of $1000$ source records
\citep{shi2025tabdiff}.
The holdout is excluded from fitting
both generators
and from training the downstream classifiers.
External predictive utility is evaluated using
extremely randomized trees (XT), bagged classification
and regression trees (bagged CART), and XGBoost (XGB),
as in Section~\ref{subsec:external_utility_results}.
Each classifier is trained on the common real training
reference and separately on the synthetic datasets,
then evaluated on the same real holdout.
The comparison includes one available TabDiff release
of $9000$ records and the fixed synthetic dataset of
the same size evaluated in
Section~\ref{subsec:external_utility_results}.
For the proposed framework, predictive performance
is summarized across ten classifier training runs
on that same dataset.
The ten additional targets in
Section~\ref{subsec:landing_stability} support
a separate stability assessment.
The comparison therefore concerns predictive performance
on a common test set, rather than variability across
repeated releases of both generators.
Table~\ref{tab:tabdiff_classification} shows decreases
in accuracy, balanced accuracy and macro-F1 for all
TabDiff-trained classifiers.
\begin{table}[htbp]
\caption{Classification performance for the TabDiff release.}
\label{tab:tabdiff_classification}
\centering
\begin{tabular}{llrrr}
\toprule
Model & Metric & Real & TabDiff & Difference \\
\midrule
XT & Accuracy & 0.4610 & 0.3120 & -0.1490 \\
XT & Balanced accuracy & 0.4597 & 0.3109 & -0.1488 \\
XT & Macro-F1 & 0.4587 & 0.3069 & -0.1519 \\
\midrule
Bagged CART & Accuracy & 0.4920 & 0.3500 & -0.1420 \\
Bagged CART & Balanced accuracy & 0.4891 & 0.3465 & -0.1426 \\
Bagged CART & Macro-F1 & 0.4885 & 0.3405 & -0.1480 \\
\midrule
XGB & Accuracy & 0.5100 & 0.4180 & -0.0920 \\
XGB & Balanced accuracy & 0.5089 & 0.4133 & -0.0956 \\
XGB & Macro-F1 & 0.5078 & 0.4005 & -0.1072 \\
\bottomrule
\end{tabular}
\end{table}
Baseline-relative accuracy losses range from $9.2$
to $14.9$ percentage points for TabDiff, whereas mean
changes are positive for the proposed framework
(Table~\ref{tab:generator_utility_gaps}).
Under these protocols, the latter retains classification
accuracy more closely relative to its real-data baselines.
\begin{table}[htbp]
\caption{Synthetic-minus-real utility differences relative to each generator's own baseline. Proposed-method differences use mean performance across ten classifier training runs on the same synthetic dataset; TabDiff differences concern one release. Positive accuracy differences and negative log-loss or Brier differences are favourable.}
\label{tab:generator_utility_gaps}
\centering
\begin{tabular}{llrr}
\toprule
Model & Metric & Proposed framework & TabDiff \\
\midrule
XT & Accuracy & 0.0246 & -0.1490 \\
XT & Log-loss & 0.1221 & 0.1154 \\
XT & Brier score & 0.0444 & 0.0737 \\
\midrule
Bagged CART & Accuracy & 0.0105 & -0.1420 \\
Bagged CART & Log-loss & 0.1707 & 0.0721 \\
Bagged CART & Brier score & 0.0584 & 0.0498 \\
\midrule
XGB & Accuracy & 0.0050 & -0.0920 \\
XGB & Log-loss & 0.0519 & 0.0899 \\
XGB & Brier score & 0.0253 & 0.0537 \\
\bottomrule
\end{tabular}
\end{table}
TabDiff log-loss values are $1.4014$, $1.3535$ and
$1.3271$ for XT, bagged CART and XGB; corresponding
Brier scores are $0.7592$, $0.7320$ and $0.7188$.
Both generators worsen probability-based scores relative
to their own baselines.
The proposed framework shows smaller increases for
both XGB scores and XT Brier, while TabDiff shows
smaller increases for both bagged-CART scores and
XT log-loss.
The comparison therefore does not establish uniform
superiority across predictive criteria.
Historical diagnostics for the TabDiff release report
mean marginal TVD $0.0410$, maximum marginal TVD
$0.1649$ and $18$ duplicate records.
These provide descriptive context rather than a harmonized
comparison of the complete fidelity objective or
the repeated MIA protocol used here.


\section{Sensitivity Analysis}
\label{sec:sensitivity_analysis}

\begin{figure}[htbp]
\centering
\includegraphics[width=\textwidth]{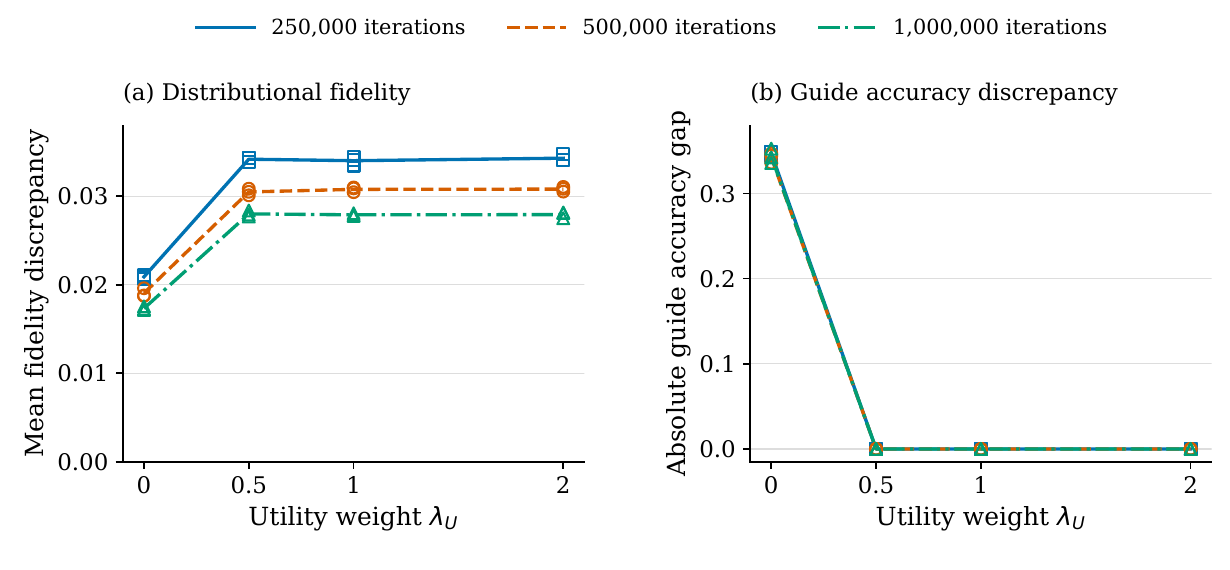}
\caption{Sensitivity of target selection outcomes to the utility weight and optimization budget. The left panel reports mean fidelity discrepancy and the right panel the absolute accuracy gap relative to the fixed random-forest guide. Lines connect means across three matched seeds; open symbols represent individual executions. The pool and endpoint temperatures are fixed. Lower values are preferable in both panels.}
\label{fig:landing_sensitivity}
\end{figure}

The sensitivity analysis examines how the balance between fidelity and utility, together with the optimization budget, affects the final synthetic target. The candidate pool and all fitted models are held fixed, so that changes in the results can be attributed to the target selection configuration and its stochastic execution. The analysis complements the fixed-configuration stability experiment reported in Section~\ref{subsec:landing_stability}.
The target selection objective is written as
\begin{equation}
H_{\lambda_U}(T)
=
F(T)
+
\lambda_U
\left|
\operatorname{Acc}_{RF}(T)
-
\operatorname{Acc}_{RF}(\mathcal{D}_R)
\right|,
\label{eq:sensitivity_objective}
\end{equation}
where $F(T)$ is the mean of the twenty fidelity components and the accuracy term is evaluated using the fixed random-forest guide. The fidelity coefficient remains equal to one. Four utility weights, $\lambda_U\in\{0,0.5,1,2\}$, are combined with three optimization budgets, $L\in\{250000,500000,1000000\}$. Each of the twelve configurations is evaluated with three matched seeds, yielding $36$ executions. The case $\lambda_U=0$ removes the utility contribution and provides a fidelity-only comparison.
For each seed, the initial target and the random-number initialization of the search are shared across configurations. The pool contains $110000$ records and each target contains $9000$ records. The initial and final temperatures remain fixed at approximately $1.59634\times10^{-4}$ and $1.59634\times10^{-7}$, respectively. Each budget uses a complete geometric cooling schedule between these endpoints; shorter runs are therefore not prefixes of the longest run. No generator or guide is refitted, and the external holdout is not used. The reported values refer to the final categorical target before numerical reconstruction.
Table~\ref{tab:landing_sensitivity} reports the mean fidelity discrepancy across the three seeds, together with its standard deviation. Increasing the optimization budget reduces the mean discrepancy for every utility weight. For the adopted configuration, $\lambda_U=1$, the mean decreases from $0.03401$ at $250000$ iterations to $0.02790$ at $1000000$ iterations, corresponding to a reduction of approximately $18\%$. The improvement from $500000$ to $1000000$ iterations is approximately $9\%$, indicating a further benefit from the longer search within the evaluated range.
\begin{table}[htbp]
\caption{Mean fidelity discrepancy across three matched seeds for each target selection configuration. Standard deviations are reported in parentheses. Lower values indicate closer distributional agreement.}
\label{tab:landing_sensitivity}
\centering
\begin{tabular}{lccc}
\toprule
Utility weight & $250000$ iterations & $500000$ iterations & $1000000$ iterations \\
\midrule
$\lambda_U=0$
& $0.02089\;(0.00018)$
& $0.01904\;(0.00049)$
& $0.01732\;(0.00021)$ \\
$\lambda_U=0.5$
& $0.03418\;(0.00027)$
& $0.03048\;(0.00037)$
& $0.02800\;(0.00035)$ \\
$\lambda_U=1$
& $0.03401\;(0.00052)$
& $0.03077\;(0.00028)$
& $0.02790\;(0.00016)$ \\
$\lambda_U=2$
& $0.03430\;(0.00038)$
& $0.03080\;(0.00027)$
& $0.02792\;(0.00036)$ \\
\bottomrule
\end{tabular}
\end{table}
Removing the utility term produces the lowest fidelity discrepancies, but leaves a substantial accuracy gap relative to the guide. For $\lambda_U=0$, the mean absolute gap is $0.34726$, $0.34252$ and $0.34356$ at the three successive budgets. Thus, increasing the number of iterations improves the selected distributional summaries without recovering the guide accuracy. At one million iterations, the mean synthetic guide accuracy is approximately $0.40689$, compared with the real-reference value of $0.75044$. This result shows that the fidelity components considered here do not, by themselves, determine the predictive behaviour measured by the guide.
In contrast, all $27$ executions with a positive utility weight attain a zero final guide accuracy gap. The final fidelity discrepancies are also similar across these weights. At one million iterations, their means range from $0.02790$ to $0.02800$, a difference smaller than the observed within-configuration standard deviations. The experiment therefore provides no clear evidence that doubling or halving the unit utility coefficient yields a systematic improvement in final fidelity.
Figure~\ref{fig:landing_sensitivity} displays the two objective dimensions separately. Although the positive weights lead to similar final outcomes, they can influence the search trajectory and the speed at which guide accuracy is attained.
These findings support the adopted choice of unit coefficients and one million iterations as a simple configuration with stable outcomes over the positive weights examined. Total energy values are not compared across weights because changing $\lambda_U$ changes the objective itself.
The conclusions are conditional on the available pool, fixed temperature endpoints and evaluated parameter range. Three seeds provide descriptive replication rather than evidence of universal parameter robustness, and the experiment does not establish a global optimum or the absence of further improvements beyond one million iterations. Moreover, the utility quantity examined here is the optimization guide accuracy; external predictive performance remains assessed separately through the three classifiers reported in Section~\ref{subsec:external_utility_results}.


\section{Discussion}
\label{sec:discussion}

The methodological contribution lies in distinguishing
record-level generation criteria from dataset-level selection
objectives within an open-box formulation.
Plausibility, privacy and diversity govern exploration,
whereas fidelity and classification utility govern selection.
This separation addresses two distinct requirements:
plausible records need not constitute a representative dataset,
and distributional agreement need not preserve the information
required for a specific predictive task.
The originality resides in the integration of established
learning, sampling and optimization methods through explicit
criteria whose roles and trade-offs can be examined.
The Bayesian network provides the connection between learned
dependencies and energy-driven exploration.
Its distribution determines proposal frequencies and the
baseline probability structure modified by the pool construction penalties.
Cancellation in the Hastings ratio follows from this dual role
and does not eliminate its influence on generation.
However, the evolving diversity archive changes the target:
detailed balance conditional on a fixed archive does not
establish convergence of the complete adaptive procedure.
Target selection operates within the support supplied by pool construction.
It can alter the representation of available configurations
but cannot recover configurations absent from the pool.
Stability across repeated selections is consequently conditional
on that pool and does not establish exhaustive exploration
or global optimality.
The sensitivity analysis informs the operational balance
between fidelity and utility, while remaining conditional on
the evaluated weights, search budgets and temperature schedule.
The objective determines which aspects of the source data
are preserved.
Averaging fidelity components permits compensation between
distributions, while matching guide accuracy constrains
classification decisions without controlling the complete
conditional response distribution.
The distinction between retained classification performance
and weaker probability-based scores is therefore consequential:
synthetic utility is specific to the analytical task.
Agreement with the optimization guide cannot establish
interchangeability with reference data for other analyses.
The comparison with the diffusion-based generator must also
be interpreted within its experimental design.  
Baseline-relative differences provide a descriptive
comparison of the evaluated releases under their
respective classifier training protocols.
The experiment does not assess variability across
repeated releases of either generator and therefore
does not establish a general ranking of generative methods. 
Selection can change configuration frequencies
and release-level risk even when every selected record
satisfies pool construction admissibility.
Assessment must therefore address the reconstructed release
and the information available to the evaluated attacks.
Because the source dataset is artificially constructed,
membership results concern disclosure of inclusion in the
experimental training reference, rather than demonstrated
protection of actual individuals.
Discretization introduces a further limitation.
It permits a common representation of mixed-type records
but removes within-interval numerical information.
Bin-preserving reconstruction retains the categorical
relationships assessed during target selection without necessarily
recovering numerical dependencies within those intervals.
Its adequacy must therefore be judged against the resolution
required by the intended analysis.
The empirical evidence remains conditional on the constructed
dataset, fitted models and generated pool.
Assessment across independently fitted generators and datasets
with different dependency structures would clarify how
discretization, pool composition and pool construction penalties affect
the attainable fidelity--utility balance.
For official statistics, the framework offers an explicit
connection between generation criteria, diagnostic evidence
and the analytical uses supported by a synthetic release.


\section{Conclusions}
\label{sec:conclusion}
This study formulates synthetic microdata generation as structured
probabilistic exploration followed by dataset-level optimization.
Its contribution is a coherent allocation of plausibility, privacy
and diversity to candidate generation, and fidelity and
classification utility to target selection. The resulting open-box
architecture makes competing requirements explicit and their
effects open to empirical assessment.
The approach treats synthetic-data quality as a set of properties
to be specified and evaluated for the intended use, rather than
inferred from generative fit alone. Its methodological value lies
in connecting the construction of a synthetic release with an
examination of the statistical information it preserves and the
disclosure risks it may retain.


\end{document}